\documentclass[conference,flushend]{iaria} 

\IEEEoverridecommandlockouts
\usepackage{amsmath,amssymb,amsfonts}
\usepackage{algorithmic}
\usepackage{graphicx}
\usepackage{textcomp}
\usepackage{xcolor}
\usepackage{tikz}
\usepackage{balance}

\allowdisplaybreaks
\DeclareRobustCommand\sampleline[1]{%
  \tikz\draw[#1] (0,0) (0,\the\dimexpr\fontdimen22\textfont2\relax)
  -- (2em,\the\dimexpr\fontdimen22\textfont2\relax);%
}

\title{Reinforcement Learning Based Goodput Maximization with Quantized Feedback \\ in URLLC}
\author{
  \IEEEauthorblockN{%
    Hasan Basri Celebi$^{1,2}$ \,\orcidlink{0000-0003-1639-3702}  and  Mikael Skoglund$^1$\,\orcidlink{0000-0002-7926-5081}}
  \IEEEauthorblockA{%
    \textit{$^1$KTH Royal Institute of Technology, Stockholm, Sweden} 
    \\
    \textit{$^2$Hitachi Energy, Västerås, Sweden}
    \\
    e-mail: {\tt hasan-basri.celebi@hitachienergy.com, skoglund@kth.se}
} }

\begin{document}
\maketitle
\begin{abstract}
This paper presents a comprehensive system model for goodput maximization with quantized feedback in Ultra-Reliable Low-Latency Communication (URLLC), focusing on dynamic channel conditions and feedback schemes. The study investigates a communication system, where the receiver provides quantized channel state information to the transmitter. The system adapts its feedback scheme based on reinforcement learning, aiming to maximize goodput while accommodating varying channel statistics. We introduce a novel Rician-$K$ factor estimation technique to enable the communication system to optimize the feedback scheme. This dynamic approach increases the overall performance, making it well-suited for practical URLLC applications where channel statistics vary over time.
\end{abstract}


\section{Introduction}

Ultra-Reliable Low-Latency Communication (URLLC) systems are facing the challenge of achieving reliability with maximum data transmission rates while dynamically responding to fluctuating channel conditions. In this context, goodput, which represents the rate of successful information transmission, is a key metric for evaluating overall system performance. Optimizing goodput emphasizes the significance of feedback mechanisms. These mechanisms let the transmitter adapt its transmission strategies efficiently \cite{Celebi_Thesis}.

\subsection{Related Work}

Various research explored scenarios where partial Channel State Information (CSI) is available, aiming to reduce system overhead compared to full feedback approaches. In \cite{Etemadi_Joint}, a more systematic feedback approach is explored, focusing on quantized CSI. Kim et al. \cite{Kim_On_the_Expected} investigates wireless communication systems with partial CSI transmitted over an error-free quantized feedback channel in the asymptotic regime, proposing an adaptive feedback scheme to maximize goodput. Recently, advancements in finite blocklength regime for low-latency communication applications have been studied in \cite{Celebi_Goodput}.

On the other hand, significant advancements in URLLC with feedback systems have been highlighted in recent years. \cite{Trung_Feedback} addresses URLLC downlink transmission quality challenges ensuring reliability and flexibility in feedback transmission. Authors in \cite{Almarshed_Deep} introduce Deep-HARQ, an AI-driven algorithm optimizing the interface design for URLLC, significantly reducing link latency. Furthermore, enhancements in downlink link adaptation for URLLC to improve the channel quality while enhancing system-level performance are presented in \cite{Pocovi_Channel}. These contributions mark significant advancements in efficient and reliable URLLC systems with feedback mechanisms \cite{Celebi_Multi}.


\subsection{Motivation and Contributions }

In this study, we assume a quasi-static fading channel where the channel coefficient $h$ remains constant during transmission but varies over different codewords. One of the primary performance metrics in quasi-static fading channels is the system's overall goodput, which can be assessed by the expected rate achieved across a substantial number of packet transmissions with varying transmission rates. This scenario requires a feedback mechanism to transmit the current CSI to the transmitter. Therefore, we propose a system model that investigates the optimum quantized feedback scheme with the purpose of maximizing the overall goodput of the communication system. To extend the existing research, it is also assumed that the channel statistics vary over time due to factors such as mobility and scattering. For this purpose, a Rician distributed channel model is taken into account with an unknown shape factor, which will be defined in the next section.

The foundation of the proposed study lies in a two-part system. The contributions in the current paper can be listed as
\begin{itemize}
    \item First, we introduce a novel technique for estimating the Rician-$K$ factor, which characterizes the channel's shape factor. This estimate serves as a key input for the second part, where Reinforcement Learning (RL)-based strategies for quantized feedback scheme selection are applied.
    \item In the second part, we propose a novel RL-based search algorithm to design an adaptive feedback scheme. RL offers a flexible approach to learning and adapting to varying communication environments.
\end{itemize}
Our approach enables transceivers to dynamically adapt feedback strategies to current channel conditions, thereby maximizing goodput. This model offers a practical solution for dynamic channel adaptation which is crucial for evolving wireless technologies and the increasing demand in next-generation industrial communications for URLLC in the upcoming beyond-5G era.

\section{System Model}

We consider the discrete-time complex baseband wireless communication system in which the transmitter transmits a codeword over a quasi-static fading channel, where the complex-valued channel coefficient $h$ is an independent and identically distributed (i.i.d.) random variable according to some distribution but remains constant over the codeword transmission. For the sake of the focus of the study, it is assumed that the receiver has perfect knowledge of $h$ and transmits this information back to the transmitter via an error-free quantized feedback channel. 

In such a communication environment the received signal $\boldsymbol{y}$ can be expressed as
\begin{equation}
    \boldsymbol{y} = h\boldsymbol{x} + \boldsymbol{z},    
\end{equation}
where $\boldsymbol{x}$ and $\boldsymbol{z}$ represent the transmitted codeword and complex Gaussian noise vector where the samples are i.i.d. and $z_i \sim CN(0,1)$, where $z_i$ represents the $i$th component of $\boldsymbol{z}$. 

Let $\gamma$ denote the i.i.d.~channel magnitude which is determined as $\gamma = |h|$, where $\gamma$ can be defined as a continuous random variable with its corresponding Probability Density Function (PDF), $p(\gamma)$, and Cumulative Distribution Function (CDF), $P(\gamma)$. In this study, it is assumed that both $p(\gamma)$ and $P(\gamma)$ are continuous and  $p(\gamma) \geq 0$ over $0 \leq \gamma \leq \infty$.

 \subsection{Feedback Channel}

It is considered that the receiver divides the positive real line into $\Lambda$ number of quantization regions and applies a deterministic index mapping on the channel magnitude $\gamma$
\begin{equation}
    L(\gamma) = l ~ \text{for} ~ \gamma \in [\lambda_l, \lambda_{l+1}), 
\end{equation}
where $l = 0, 1, \cdots, \Lambda - 1 $ and $\lambda_0 = 0$ and $\lambda_{\Lambda} = \infty$. Afterward, the selected index, $l$, is transmitted back to the transmitter over the error-free feedback channel. Therefore, CSI is partially known to the transmitter.

After receiving the partial CSI, the transmitter selects a transmission rate, described as $r_l$, which can be defined as the selected rate for the $l$th quantized region, with the mission of maximizing the goodput of the communication system, which is the maximization of the overall correctly received information rate. For instance, the goodput of a communication system with constant transmission rate $r$ and error rate $\epsilon$ is
\begin{equation}
    G = r(1-\epsilon).
\end{equation}

\subsection{Problem Definition}

The instantaneous channel capacity, for a given channel magnitude $\gamma$ with SNR $\mathcal{P}$, is 
\begin{equation}
    C(\gamma) = \log (1+\gamma^2 \mathcal{P}).
\end{equation}
Suppose $\Lambda = \infty$, which represents perfect CSI at the transmitter, the maximum achievable goodput is the ergodic capacity since it is possible to match the transmission rate to $C(\gamma)$. Thus,
\begin{equation}
    G_{\Lambda = \infty} = \int^{\infty}_{0} p(\gamma) C(\gamma) \text{d}\gamma.
    \label{eq_ergodic_capacity}
\end{equation}
On the other hand, if $\Lambda = 1$, which means no CSI at the transmitter, the maximum achievable goodput can be found by solving the following optimization problem
\begin{equation}
    G_{\Lambda = 1} = \max_{r \geq 0} \int^{\infty}_{\sqrt{\frac{2^r-1}{\mathcal{P}}}} r p(\gamma) \text{d}\gamma.
    \label{eq_no_csi_capacity}
\end{equation}
When $\Lambda \in [2, \infty)$, determining the maximum achievable goodput value becomes a challenging task. Consequently, the objective becomes identifying the optimal $\lambda_l$ and $r_l$ configurations that maximize the long-term goodput. 

\begin{figure}
    \centering
    \includegraphics[width=.50\textwidth]{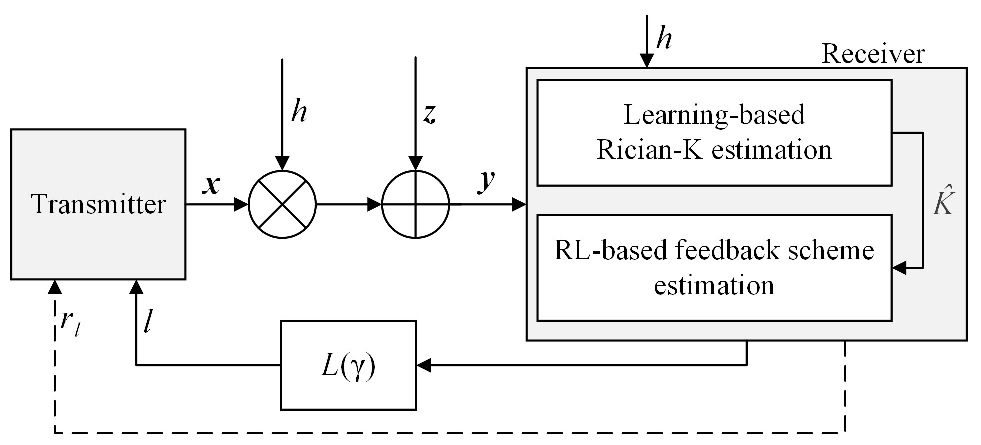}
    \caption{The proposed system model}
    \label{fig_system_model}
\end{figure}

Suppose $\gamma^r_l$ is the reconstruction point of the $l$th quantization region and the transmitter selects $r^r_l = C(\gamma^r_l)$ as the transmission rate. For a given channel realization $\gamma$, where $\lambda_l \leq \gamma < \lambda_{l+1}$, we know that as long as 
\begin{equation}
    r^r_l \leq \log(1+\gamma^2 \mathcal{P}) = C(\gamma), 
\end{equation}
error-free transmission is possible. Otherwise, the communication is in outage. Thus, once the $\gamma^r_l$s and $r^r_l$s for $l=1,2,\cdots, \Lambda$ are estimated, the overall outage probability of such a communication system is
\begin{equation}
    \sum_{l=0}^{\Lambda-1} (P(\gamma^r_l) - P(\lambda_l)) .
\end{equation}
Therefore, the optimum selections of $\lambda_l$s and $r_l$s can be found by solving the following optimization problem
\begin{subequations}
\begin{align}
    \max_{r^r_l, \lambda_l, \gamma^r_l} & ~~ \sum_{l=0}^{\Lambda-1} r^r_l \big( P(\lambda_{l+1}) - P(\gamma^r_l) \big)
    \\
    \text{s.t.} & ~~ 0 < \lambda_l \leq \gamma^r_l < \lambda_{l+1} < \infty .
\end{align}
\label{eq_goodput_opt_problem}
\end{subequations}

This problem has been studied in \cite{Kim_On_the_Expected, Makki_On_Hybrid}, and \cite{Celebi_Goodput}, and it was shown that optimum $r^r_l$s can be achieved by setting $r^r_l = C(\lambda_l)$ and quantization levels, $\lambda^r_l$ for $l = 1, 2, \cdots, \Lambda-1$, can be found by solving the following equation with an iterative algorithm
\begin{equation}
    P(\lambda^r_{l+1}) = P(\lambda^r_l) + \left( \frac{1}{\mathcal{P}}  + \lambda^r_l\right) p(\lambda^r_l) \log\left( \frac{1+\lambda^r_l \mathcal{P}}{1+\lambda^r_{l-1} \mathcal{P}} \right) .
\end{equation} 

\subsection{Varying Channel Statistics}

Notice that the channel statistics are assumed to be fixed in the optimization problem above. In this paper, we extend these results by \emph{letting the channel statistics vary over time}. 

It is assumed that $h \sim CN(\mu, \sigma^2)$, with slowly varying $\mu$ over time\footnote{In our analyses, we assumed that $\mu$ remains constant for $200$ channel realizations and then changes to a new value.}. Since $h \sim CN(\mu, \sigma^2)$, the channel magnitude, $\gamma$, is Rician distributed random variable with varying $K$-factor, which represents the shape parameter of the distribution and can be defined as the power ratio of the line-of-sight signal power to the remaining multipath and is expressed as \cite{Abdi__On_The} 
\begin{equation}
    K = \mu^2/\sigma^2 .
\end{equation}
This assumption is broader and more realistic for many real-world applications where the channel statistics change due to mobility, scattering, etc.

For this purpose, we divide the proposed method into two parts: $i$) Rician-$K$ factor estimation and $ii$) RL-based quantized feedback scheme selection. In the first part, we introduce a novel technique for estimating the Rician-$K$ factor. The estimate obtained in this initial phase serves as an input for the subsequent part, where an RL-based strategy is employed to dynamically select and update the feedback scheme to optimize the overall goodput of the communication system, aligning it with the current channel statistics. The proposed system model is shown in Fig. \ref{fig_system_model}.

In Fig. \ref{fig_system_model}, it is worth highlighting the presence of two distinct feedback channels. The first feedback channel, depicted with a solid line, serves as the quantized feedback channel and is utilized in each transmission of a codeword. In contrast, the second feedback channel, indicated by a dashed line, plays a unique role during the training phase, specifically for transmitting updated transmission rates associated with quantization level $l$. It is important to emphasize that this secondary feedback channel is not employed in every subsequent transmission; rather, its usage is triggered by the decision made by the RL-based feedback scheme to update $r_l$, as will be discussed in Section IV. In this way, the transmitter's knowledge is restricted to the selected transmission rate $r_l$ for each $l$, which minimizes the need for additional data transmission during training since all learning algorithms are implemented at the receiver. This also contributes to reducing the overall computational load on transceivers, which is a crucial factor for mitigating the latency due to computational processes \cite{Celebi_Low_Latency} and \cite{Celebi_Latency}. 

\begin{figure*}
    \centering
    \includegraphics[width=1\textwidth, clip=true, trim = 20mm 5mm 20mm 10mm]{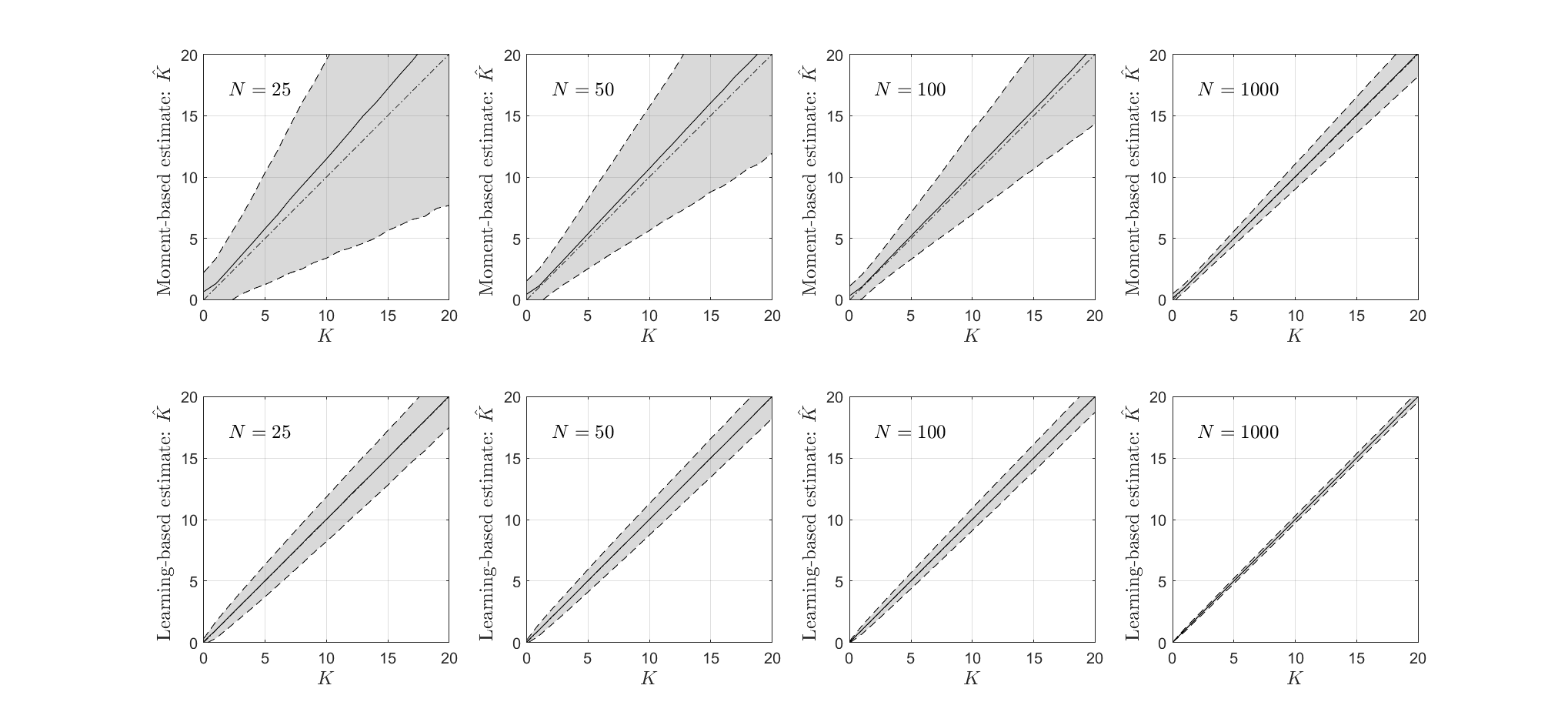}
    \caption{Sample mean and sample confidence region of the two estimators, namely moment-based and learning-based estimators. Results have been depicted for $N=\{25, 50, 100, 1000\}$. (\sampleline{}) Sample mean. (\sampleline{dashed}) Upper and lower limits of the confidence region. (\sampleline{dash pattern=on .7em off .2em on .05em off .2em}) Reference line.}
    \label{fig_K_est_results}
\end{figure*}

\section{Learning-Based Estimator for Rician-K Factor}

Rician-$K$ factor estimation is a well-studied topic in the literature. Moment-based and maximum-likelihood estimators have been presented in \cite{Abdi__On_The, Nicolas_A_New}. Here, we first introduce the moment-based estimators and then present our findings. 

The PDF of the channel magnitude $\gamma$ is given by
\begin{equation}
    p(\gamma) = \frac{2\gamma}{\sigma^2} \exp\Big( \frac{-(\gamma^2 + \mu^2)}{\sigma^2} \Big) \text{I}_0\Big( \frac{2\gamma\mu}{\sigma^2} \Big),
\end{equation}
where $\text{I}_z(\cdot)$ is the modified Bessel function of the first kind with order $z$. By using the first few raw moments of Rician distribution \cite{Rice_Mathematical}, the following estimators can be found after some straightforward mathematical steps
\begin{align}
    \frac{m_1}{\sqrt{m_2}} &= \frac{1}{2} \sqrt{\frac{\pi}{K+1}} L_{\frac{1}{2}}(K), \label{eq_first_mom_est}
    \\
    \frac{m_4}{m_2^2} &= 1+\frac{(2K+1)}{(K+1)^2}, 
    \label{eq_second_mom_est}
    \\
    \frac{m_6}{m_2^3} &= 6 + \frac{K^2(5K-9)}{(K+1)^3},
    \label{eq_third_mom_est}
\end{align}
where $L_{\frac{1}{2}}(K)$ is the Laguerre polynomial, defined as 
\begin{equation}
    L_{\frac{1}{2}}(K) = \exp\left(-\frac{K}{2}\right) \left( (K+1) \text{I}_0\left(\frac{K}{2}\right) + K \text{I}_1 \left(\frac{K}{2}\right) \right), 
\end{equation}
and $m_i$ represents the $i$th raw moment. By using the estimators above, an estimate of $K$ is obtained numerically by computing the empirical moments and solving the nonlinear equations presented, where a close approximation for the Laguerre polynomial can also be used as in \cite{Celebi_Training}.

In this study, we extend these results and propose a novel moment-based learning model for Rician-$K$ estimation. In this proposed method, we employ a more comprehensive set of features unlike the already presented studies in the literature \cite{Alymani_Rician}, which uses the traditional amplitude samples directly. Specifically, we extract the first ten empirical moments of the Rice-distributed random variable as input features. These moments include the empirical mean, variance, skewness, and kurtosis.

The idea behind using these moments as input features is because of their ability to capture the underlying statistical characteristics of the Rician random variable $\gamma$. This approach offers a more detailed and informative representation of the channel compared to conventional moment-based estimators, which often rely only on a few moments. Our learning-based approach is based on the eXtreme-Gradient-Boosting (XGBoost) regression model, which has gained popularity in various domains for its capability to handle complex relationships with high-performance predictions \cite{Hsu_A_Boosting}. Other learning algorithms such as linear regression, histogram-based gradient boosting regression, random forest regressor, cat-boost regressor, etc. are also investigated. We skipped their results since their overall performance was worse than XGBoost.

\subsection{Pre-processing}

Notice that the number of inputs of the proposed learning-based method does not change with the number of samples collected, which makes the proposed method easily scalable. On the other hand, identifying the number of samples, $N$, while computing the empirical raw moments becomes a significant design problem. To find the best selection, we have tested various $N$ values and saw that the best performance is obtained when the training dataset is comprised of $N=100$. However, similar performance results can be achieved with selections of $N>50$ since XGBoost uses a learning rate to control the model's parameter updates during training \cite{Nielsen_Why_Does}, which aims to prevent overfitting while maintaining a low bias and ensuring that the model generalizes well to unseen data.

\subsection{Performance Comparison}

To test the performance of the proposed method, we resorted to Monte Carlo simulations and compared the results. For this purpose, a training dataset comprising $10^5$ Rician-$K$ factors is created. This dataset contains a range of $K$ values, limited within $0 \leq K \leq 100$, and was appended with their respective empirical raw moments, which were computed from randomly generated samples, with each dataset comprising $N=100$ samples. Even though we set a constant $N$ for the training dataset, we test the performance of the learning-based estimator with various $N$ values, which are $N=\{25, 50, 10^2, 10^3\}$. On the other hand, we select the estimator formulated in \eqref{eq_first_mom_est} as the moment-based estimator due to its leading performance against \eqref{eq_second_mom_est} and \eqref{eq_third_mom_est} since it uses the advantage of the lower order moments \cite{Abdi__On_The}. 

A comparison of the results is presented in Fig. \ref{fig_K_est_results}, where the sample mean of the predicted $K$ values, $\hat{K}$, are depicted with upper and lower limits of the confidence region, which is calculated as the $\pm 2 \times$standard deviation of $\hat{K}$. Although the estimators are capable of detecting higher $K$ values, we limit the horizontal and vertical axes between $[0-20]$ to have a better look at the differences between the estimators. 

The key observation derived from Fig. \ref{fig_K_est_results} is the remarkable superiority of the learning-based estimator in comparison to the moment-based estimator across all choices of $N$. Interestingly, the learning-based estimator exhibits better performance, even at $N=25$, compared to the moment-based estimator's results at $N=10^2$. Furthermore, the learning-based estimator achieves nearly perfect estimations when $N=10^3$.

\section{Reinforcement-Learning Based Quantized Feedback Scheme Selection}

In this section, we focus on finding the optimum feedback scheme, such as selecting the best $\lambda_l$s and $r_l$s, to maximize the overall goodput of the system. As mentioned in the previous chapter, we use an RL-based search algorithm since the proposed methods in the literature cannot adapt to variable channel statistics. 

Let us first denote the optimum selections as $\lambda_l^*$ and $r_l^*$. It is shown in \cite{Kim_On_the_Expected} that the optimum goodput is achieved by assigning the $r_l^* = C(\lambda_l^*)$. Thus, one can simplify the optimization problem by omitting the $r_l$ variables. Next is the discritization process of the quantization regions. For this purpose, we define a finite number of values for $\lambda_l$s and reformulate the optimization problem in \eqref{eq_goodput_opt_problem} as
\begin{subequations}
\label{eq_goodput_opt_problem2}
\begin{align}
    \max_{\lambda_l, \gamma_l} & ~~ \sum_{l=1}^{\Lambda-1} r_l \big( P(\lambda_{l+1}) - P(\gamma_l) \big)
    \\
    \text{s.t.} & ~~ 0 \leq \lambda_l \leq \gamma_l < \lambda_{l+1} < \infty , \label{eq_const_1}
    \\
    & ~~ \lambda_l \in \mathcal{S}, 
\end{align}
\end{subequations}
where $\mathcal{S}$ represents the set of finite number of selections. With this reformulation, the solution becomes a sequential search and can be modeled as a Markov decision process and therefore can be solved with RL \cite{Hoang_Deep_RL}. 

A Markov decision process consists of four elements, such as the environment, state space $\mathcal{S}$, action space $\mathcal{A}$, and reward space $\Omega$. In more detail, at each time step $t$ (or at each iteration) the process is in state $s_t\in \mathcal{S}$ and makes a decision and chooses an action $a_t \in \mathcal{A}$. A reward $\omega_t \in \Omega$ is observed after taking the action. Thus, $\omega_t$ is received from the environment and is based on $s_t$ and $a_t$.

For the reformulated optimization problem in \eqref{eq_goodput_opt_problem2}, states, actions and rewards are designed as follows:
\begin{itemize}
\item\textit{State space} $\mathcal{S}$: We set $\lambda_l$s for $l = \{1, 2,\cdots, \Lambda - 1 \}$ as the agents of the system, except $\lambda_0$ and $\lambda_{\Lambda}$ since their locations are fixed. The set of possible selections of $\lambda$ defines $\mathcal{S}$, which is a subset of $\mathrm{R}^+$. Therefore, $s_t$ is defined to be a vector consisting of the current locations of $\lambda_l$s for $l = \{0, 1, 2,\cdots, \Lambda \}$.

\item\textit{Action space} $\mathcal{A}$: We design the RL algorithm in such a way that in every subsequent iteration only one agent, e.g. $\lambda_l$, can change its status. Thus, 
\begin{equation}
    a_t \in \{-1, 0, +1\}, 
\end{equation}
where $-1, 0,$ and $1$ represent decreasing the index of the agent in $\mathcal{S}$ by one, no change, and increasing the index of the agent in $\mathcal{S}$ by one, respectively. Here, we also apply the $\epsilon$-greedy strategy \cite{Masadeh_RL}, which can be defined as selecting the best action with probability $(1-\epsilon_t)$ and a uniformly distributed random action with probability $\epsilon_t$, which gives the search algorithm the possibility to explore the whole state space $\mathcal{S}$ without getting stuck into a local maximum. Additionally, it is important to note that any selected action $a_t$ shall not cause any contradiction with the constraint defined in \eqref{eq_const_1}.

\item\textit{Reward space} $\Omega$: The reward function is defined as the empirical mean of the goodput that is achieved with the new state $s_t$, i.e.
\begin{equation}
    \omega_t = \frac{1}{M}\sum_{m=1}^M  C(\lambda_{L(\gamma_m)}) \triangleq G^M_{\text{emp}},
    \label{eq_reward}
\end{equation}
where $M$ represents the number of transmitted blocks with state $s_t$ and yet is another design configuration of the RL model which will be discussed in the next section. 
\end{itemize}

\subsection{Q-Learning for Enhancing Goodput}

The traditional approach in reinforcement learning involves a method known as temporal difference (TD) learning. This technique blends aspects of both Monte Carlo and dynamic programming. It is similar to Monte Carlo since TD learning acquires samples directly from the environment, and it is similar to dynamic programming since it refines its estimates based on both the current and previous assessments. One of the main TD learning methods is Q-learning which can be represented as an RL methodology allowing the agent to acquire the best strategy for navigating a specific environment \cite{Jang_Q_Learning}. This requires the agent to keep track of an approximation of the anticipated long-term discounted rewards for every possible state-action combination and then make choices to maximize these rewards. 

In the Q-learning process, the agent iteratively updates its Q-table, which stores the expected cumulative rewards for each state-action pair. The optimal policy can be found by Bellman's optimality equation \cite{Jang_Q_Learning}
\begin{equation}
    Q^*(s,a) = \mathrm{E}[\omega_{t+1} + \eta \max_{a'}Q^*(s_{t+1}, a')|s_t=s, a_t=a]
    \label{eq_Q_optimum}
\end{equation}
where $\eta$ represents the discount factor which is required to bound the cumulative reward and $\max_{a'}Q^*(s_{t+1}, a')$ defines the best estimate for the next state $s_{t+1}$. \eqref{eq_Q_optimum} reveals that Q-learning requires the current state-action pair, the resulting reward, and the subsequent state. Thus, the updating process of the Q-table can be formulated as
\begin{multline}
    Q(s_t,a_t) = (1-\alpha)Q(s_t,a_t) + 
    \\
    \alpha (\omega_t + \eta \max_{a'}Q^*(s_{t+1}, a')),
    \label{eq_updating_Q_table}
\end{multline}
where $\alpha$ is the learning rate. 

The total number of possible $s_t$s in the current problem can be expressed as 
\begin{equation}
    |\mathcal{S}|! / ((\Lambda+1)!(|\mathcal{S}|-\Lambda-1)!), 
\end{equation}
where $|\cdot|$ represents the cardinality of the set. Given the potential for rapid growth of this number, constructing a comprehensive Q-table covering all possible $s_t$ states becomes impractical. To address this challenge, we adopt a strategy where the algorithm treats each agent independently, thus it only needs the creation of distinct Q-tables for each agent. Additionally, given the variability in channel characteristics, the Q-table must encompass all feasible $K$ values. 

Consequently, the Q-learning table size for each agent is determined as
\begin{equation}
    |\mathcal{K}||\mathcal{S}|,
\end{equation}
where $\mathcal{K}$ is the set of all possible Rician-$K$ factors. Notably, it remains possible to append these distinct Q-tables into a unified table with size 
\begin{equation}
    (\Lambda-1)|\mathcal{K}||\mathcal{S}| + 1.
\end{equation}

\subsection{Algorithm Design}

The proposed algorithm is implemented as follows
\begin{itemize}
    \item Set $M$, $\alpha$, $\eta$, and initialize the $\lambda_l$ values for all agents.
    \item Initialize the Q-learning table with zeros. Set the $\epsilon$ values for all agents, where $\epsilon_1 = 0.5$ in our implementation. 
    \item Start the loop by selecting an agent, i.e. $\lambda_l$, where at each iteration a different agent is selected.
    \item Select an action based on the $\epsilon$-greedy algorithm. 
    \item Update $\epsilon_t$ with respect to $t$, such that $\epsilon_{t+1} = \epsilon_t/\sqrt{t}$.
    \item After selecting $a_t$, update $\lambda_l$ and $s_t$.
    \item Send the new $r_l$ to the transmitter. 
    \item Observe $M$ number of transmissions, compute the reward $\omega_t$ according to \eqref{eq_reward}, and update the Q-table using \eqref{eq_updating_Q_table}.
\end{itemize}
It is important to note that the proposed algorithm may not yield the optimal feedback scheme, which is due to the simplification of the Q-table size, as explained earlier, but it achieves a close approximation.

\section{Performance Evaluation}

\begin{figure}[t]
    \centering
    \includegraphics[width=0.49\textwidth, clip=true, trim = 5mm 0mm 10mm 5mm]{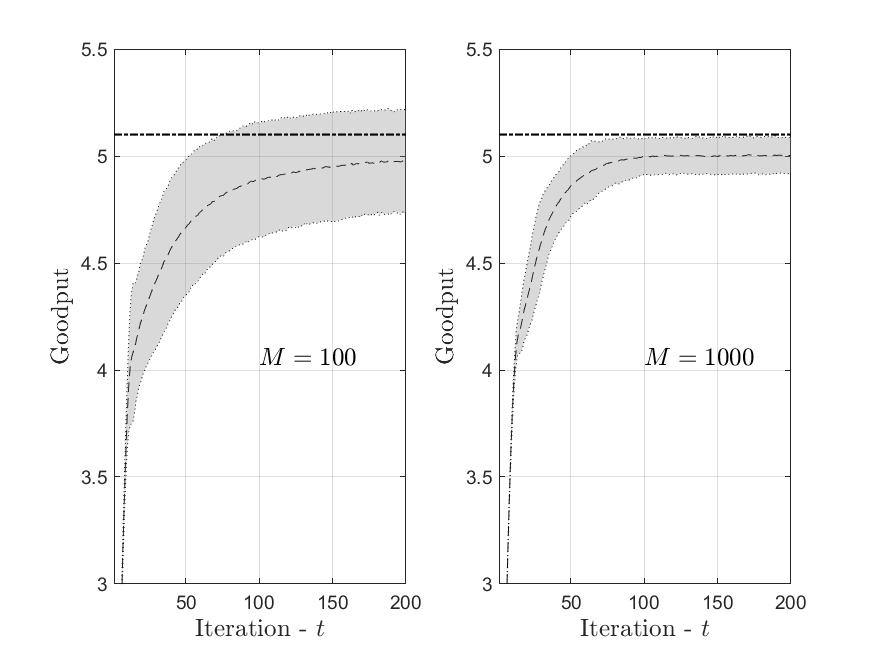}
    \caption{Mean and confidence region of $\omega_t$ with respect to iteration $t$ for $M = \{100, 1000\}$ when $\Lambda = 4$,  $K = 10$dB, and $\mathcal{P} = 20$dB. (\sampleline{dash pattern=on .7em off .2em on .05em off .2em}) The long-term average of the maximum achievable goodput. (\sampleline{dashed}) Average of $\omega_t$.}
    \label{fig_M_perf}
\end{figure}

Here, we first show the effect of $M$ on the algorithm performance. For this purpose, we implement a Monte Carlo simulation where $\Lambda = 4$,  $K = 10$dB, and $\mathcal{P} = 20$dB for $M = \{10^2, 10^3\}$ and show the average of $\omega_t$ at each iteration $t$ with the variance with grey color around the average. Results are depicted in Fig. \ref{fig_M_perf}, where the long-term maximum achievable goodput value is also plotted as a dash-dotted line. The results demonstrate that the effect of $M$ is significant for the design of the system. As can be seen from Fig. \ref{fig_M_perf}, the algorithm can reach the optimum value in both cases but faster, in terms of $t$, when $M=10^3$. However, note that higher $M$ values require more transmission at each iteration. On the other hand, in some cases, $\omega_t$ exceeds the upper limit due to its empirical nature.

\begin{figure}[t]
    \centering
    \includegraphics[width=0.49\textwidth, clip=true, trim = 5mm 0mm 10mm 5mm]{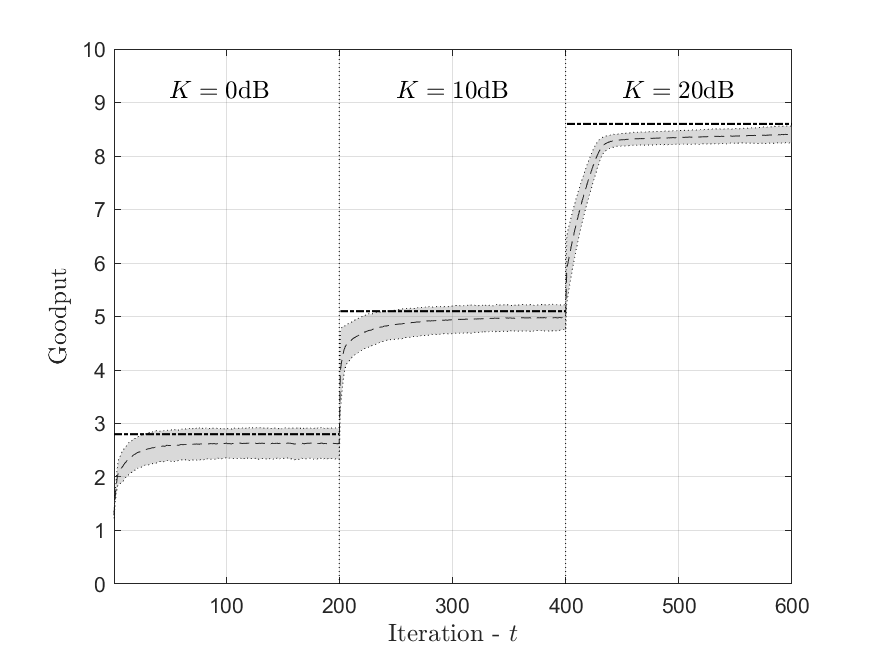}
    \caption{Performance of the proposed method with varying $K$. (\sampleline{dash pattern=on .7em off .2em on .05em off .2em}) Long-term average of the maximum achievable goodput. (\sampleline{dashed}) Average of $\omega_t$.}
    \label{fig_K_perf}
\end{figure}

Next, we focus on the overall performance of the proposed method, where we set $\Lambda = 4$, $\mathcal{P} = 20$dB, $N=10^2$, and $M=10^2$, and let $K$ change from $0$dB to $10$dB then to $20$dB. To see the overall performance, we again implement a Monte Carlo environment, from where the average $\omega_t$s at each iteration $t$ are obtained and depicted in Fig. \ref{fig_K_perf}. The long-term maximum achievable goodput values for each $K$ are also plotted with dash-dotted lines. It is possible to see that the proposed method can track the change in channel statistics and adapt its feedback scheme so that it can approach the maximum achievable goodput values in every case. It is also important to highlight that, thanks to the RL-based learning approach, once the optimal feedback scheme for a particular Rician-$K$ factor is determined, the transceivers can instantly adjust to the optimal scheme whenever the channel exhibits the same $K$ value.

\section{Conclusions}

In this study, we introduce a learning-driven system for goodput maximization with quantized feedback in wireless communication, designed to meet the requirements of URLLC. Our contributions include a novel Rician-$K$ factor estimation technique that improves the adaptability of feedback strategies to changing channel conditions. Additionally, we employed reinforcement learning (RL) to dynamically select and update feedback schemes, demonstrating the system's ability to maximize goodput under evolving channel conditions. The importance of dynamic feedback mechanisms is emphasized, which addresses the unique challenges posed by URLLC in next-generation wireless networks. Future research could extend the proposed framework to various other wireless communication scenarios.


\printbibliography
\end{document}